\documentclass[preprint]{vgtc}               % preprint

\ifpdf%                                % if we use pdflatex
  \pdfoutput=1\relax                   % create PDFs from pdfLaTeX
  \usepackage{graphicx}                % allow us to embed graphics files
  \DeclareGraphicsExtensions{.pdf,.png,.jpg,.jpeg} % for pdflatex we expect .pdf, .png, or .jpg files
\else%                                 % else we use pure latex
  \usepackage{graphicx}                % allow us to embed graphics files
  \DeclareGraphicsExtensions{.eps}     % for pure latex we expect eps files
\fi%

\graphicspath{{figures/}{pictures/}{images/}{./}} % where to search for the images

\usepackage{microtype}                 % use micro-typography (slightly more compact, better to read)
\PassOptionsToPackage{warn}{textcomp}  % to address font issues with \textrightarrow
\usepackage{textcomp}                  % use better special symbols
\usepackage{mathptmx}                  % use matching math font
\usepackage{times}                     % we use Times as the main font
\usepackage{cite}                      % needed to automatically sort the references
\usepackage{tabu}                      % only used for the table example
\usepackage{booktabs}                  % only used for the table example
\usepackage{graphicx}
\usepackage{subcaption}
\usepackage{enumitem}
\usepackage{todonotes}
\usepackage{color}
\usepackage[linkbordercolor={0 0 1}]{hyperref}
\onlineid{0}

\vgtccategory{Research}

\title{Manifesto for Putting `Chartjunk' in the Trash 2021!}

\author{Derya Akbaba\thanks{e-mail: derya@cs.utah.edu}\\ %
        \scriptsize University of Utah %
\and Jack Wilburn\thanks{e-mail: jwilburn@sci.utah.edu}\\ %
     \scriptsize University of Utah %
\and Main T. Nance\thanks{twitter: \href{https://twitter.com/SayNo2Chartjunk}{@SayNo2Chartjunk}}\\ %
     \scriptsize The Internet%
\and Miriah Meyer\thanks{e-mail: miriah.meyer@liu.se}\\ %
     \parbox{1.4in}{\scriptsize \centering Linköping University}}

\teaser{
  \centering
  \includegraphics[width=0.6\columnwidth]{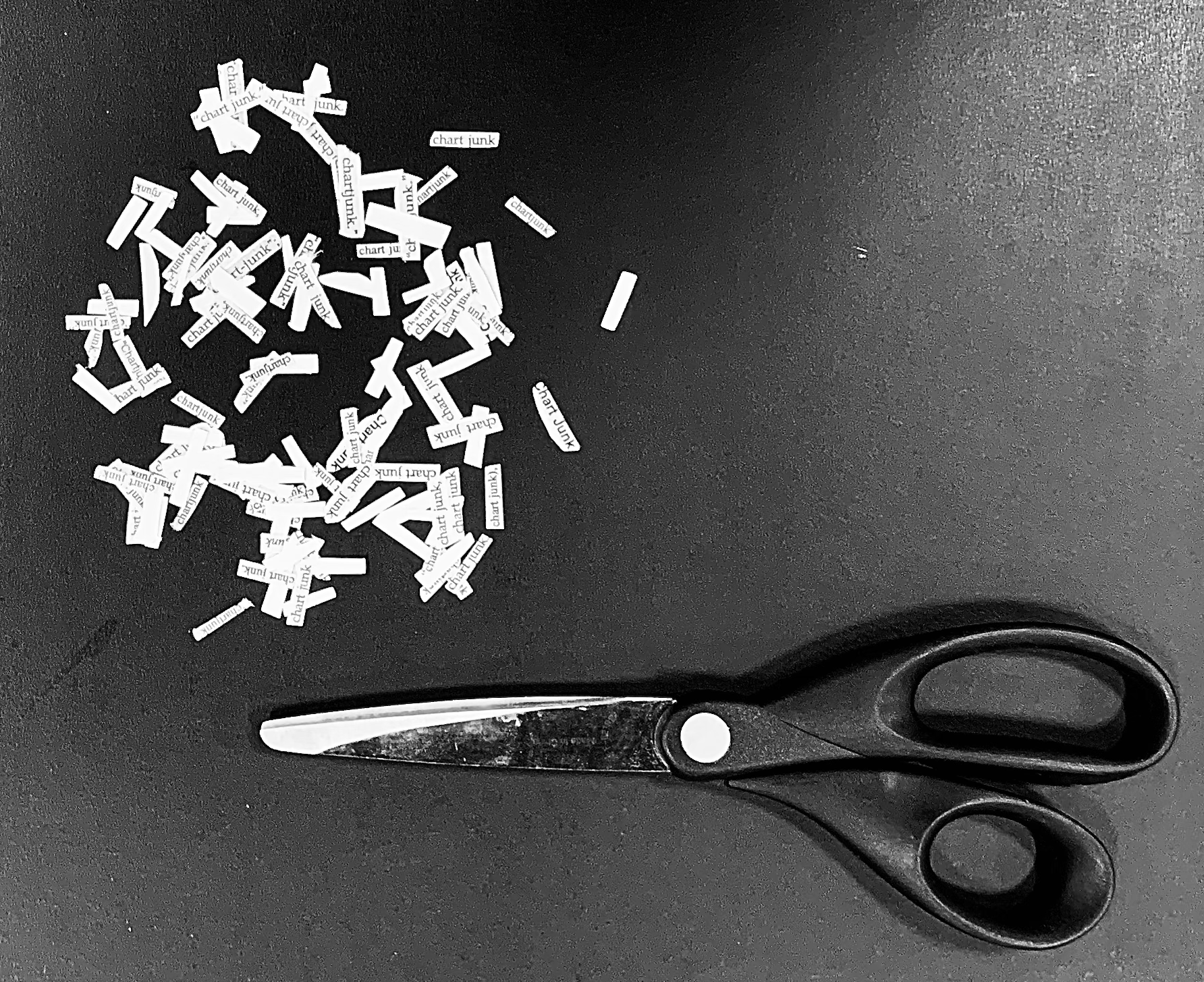}
  \caption{The remains of maintenance art. Material scraps of the term chartjunk, all destined for the trash.}
  \label{fig:teaser}
}
\preprinttext{This manuscript was  presented at alt.VIS, a workshop co-located with IEEE VIS 2021 (held virtually)}

\abstract{
In this provocation we ask the visualization research community to join us in removing chartjunk from our research lexicon. We present an etymology of chartjunk, framing its provocative origins as misaligned, and harmful, to the ways the term is currently used by visualization researchers. We call on the community to dissolve chartjunk from the ways we talk about, write about, and think about the graphical devices we design and study. As a step towards this goal we contribute a performance of maintenance through a trio of acts: editing the Wikipedia page on chartjunk, cutting out chartjunk from IEEE papers, and scanning and posting a repository of the pages with chartjunk removed to invite the community to re-imagine how we describe visualizations. This contribution blurs the boundaries between research, activism, and maintenance art, and is intended to inspire the community to join us in taking out the trash.
} % end of abstract

\begin{document}


\maketitle


\section{Introduction}

The term \textit{chartjunk} was coined by Edward Tufte in 1983 as a provocation meant to challenge what he considered to be the ``inept graphical work" of leading visualization practitioners of the time\cite[p.79]{tufte2001}. Chartjunk, according to Tufte, is graphical decoration, or ``non-data-ink", erasable and providing nothing new to the viewer \cite[p.107]{tufte2001}. Tufte's provocation stems from his embrace of minimalist design that rejects decoration and visual clutter in favor of clean lines and simple, abstract design. It is, as Kosara poignantly summarizes, a perspective that maintains ``decoration is a sin" \cite{kosara2012}.

We, the visualization research community, took Tufte's chartjunk provocation seriously and integrated the term into our courses, studies, and papers. We use it to describe a broad range of graphical elements, from redundant data encodings to annotations, iconography, and imagery \cite{bateman2010,borkin2013,parsons2020,wu2021}. The provocative stance of chartjunk, however, has undermined our efforts to precisely understand and characterize the role of non-data-ink in a viewer's experience of reading a chart. We contradict ourselves when we describe productive cognitive effects as stemming from something akin to trash, fumbling with other value-laden words like embellishment\cite{bateman2010} or ornamentation\cite{hullman2011}; and we discount the situated and personal aspects of visualization design when we call efforts to engage viewers rubbish\cite{parsons2020}. When we integrated chartjunk into our research lexicon it inevitably carried with it all the judgments and controversies associated with its origins. 

In this, our provocation, we suggest that it is time to take out the trash: that the visualization research community put an end to the use of the word chartjunk when describing purposeful, visual design elements. We call on the community to acknowledge the value-laden origins of chartjunk and to begin the work of developing and adopting more precise language. 

Towards this goal we report on performance of maintenance\footnote{Artifacts from this performance can be found online: \url{https://jackwilb.github.io/chart-junk/}}\footnote{The original link had been https://chartjunk.art/, but sadly paying for a domain requires maintenance in the form of money. So on 9 Sept. 2024, we removed the fun, cool custom domain.} that enacts our own commitment to the work of maintaining a thoughtful research lexicon. 
The performance consists of three actions that are representative of taking out our community's trash, inspired by the maintenance art of Mierle Laderman Ukeles. Ukeles' performance art blurred the lines of maintenance and art, drawing attention to all the work it takes to upkeep a home and a city. Our actions work to upkeep our research lexicon and resulted in a triptych of artifacts that serve as traces of our work:
\begin{itemize}[nosep]
\item an edit of the chartjunk \href{https://en.wikipedia.org/wiki/Chartjunk}{Wikipedia page} to provide a richer etymology of the term -- an act of maintenance activism
\item   a video of the first author cutting out the term chartjunk from the catalog of IEEE papers -- a demonstration of maintenance as art, and art as maintenance
\item  a scanned repository of the remains of the cut papers -- an invitation for the community to brainstorm new, precise terminology
\end{itemize}
These actions and their resulting artifacts are just a first step towards putting chartjunk in the trash. We call on the community to continue the maintenance of ensuring that our research lexicon is an accurate reflection of our values and our goals.

\section{Etymology of chartjunk}
\label{sect:facts}

In this etymology we differentiate between the provocative origins of chartjunk and its evolution as a research term. In this differentiation we draw attention to the mismatch between the work the term was intended to do versus the ways the visualization research community uses it. The mismatch leads to challenges in precisely and consistently describing various graphical devices, and troubling implicit valuation of one design aesthetic over others. Our telling of the origins and evolution of chartjunk is itself a provocation, and we end this section with a call to the community to dissolve and replace chartjunk in our research lexicon. 

\subsection{Origin}

In 1983 Tufte self-published the first of his well-known visualization books, \textit{The Visual Display of Quantitative Information}\cite{tufte2001}. The book was based on ideas and materials he developed for a seminar at Princeton he taught with renowned statistician John Tukey. At its core, Tufte's first book is a political manifesto for a minimalist, statistical view of what makes for good visualization design, born from his training as both a statistician and a political scientist. He argued that ``graphics reveal data” when they are designed with ``graphical integrity”, and called for creating visualizations that ``tell the truth about data” through unbiased, and unadorned charts. His embrace of minimalist design was rooted in a commitment to the objective and neutral values of science, reflected in his desire for visualizations to serve as ``instruments for reasoning about quantitative information”\cite[p13,53]{tufte2001}. Scholars have noted, however, that conventions of minimalist visualization design are also rhetorical devices that can be just as biased \cite{bateman2010, kennedy2016work, correll2014, kostelnick2008}. 

In his book, Tufte took direct aim at the leading visualization practitioners at the time when he posed: ``Why do artists draw graphics that lie? Why do the world's major newspapers and magazines publish them?" He specifically called out the ``inept graphics" of the then newly appointed first visualization specialist at \textit{Time Magazine}, Nigel Holmes, as stemming from his lack of formal, statistical training: ``Nearly all those who produce graphics for mass publication are trained exclusively in the fine arts and have had little experience with the analysis of data. Such experiences are essential for achieving precision and grace in the presence of statistics... Those who get ahead are those who beautified data, never mind statistical integrity"\cite[p79]{tufte2001}. 

In his follow-up book, \textit{Envisioning Information}\cite{tufte1990}, Tufte further critiqued the work of Holmes with a scathing criticism of his \textit{Diamonds} chart: ``Consider this unsavory exhibit at right -- chockablock with clich\'e and stereotype, coarse humor, and
a content-empty third dimension... Credibility vanishes in clouds of chartjunk; who would trust a chart that looks like a video game?"\cite[p34-35]{tufte1990}.  A 1992 \textit{New York Times} article describes the response from Holmes: ``\textit{Time's} Nigel Holmes, creator of the diamonds graph, was understandably irked when Tufte criticized it. Holmes admits his work has sometimes been exaggerated, but feels that Tufte, in his insistence on absolute mathematical fidelity, remains trapped in `the world of academia' and insensitive to `the world of commerce,' with its need to grab an audience'' \cite{nyt1992}. These critiques set up a debate between Tufte and Holmes, pitting statistical and designerly approaches to visualization design against each other \cite{few2011,bateman2010,parsons2020}.

To address ``deceptive displays" stemming from what Tufte viewed as lax visualization standards, he proposed several guidelines to help ensure graphical integrity. One of the most well-known is the elimination of what Tufte provocatively calls \textit{chartjunk}. Chartjunk is any graphical element that ``does not tell the viewer anything new". Equating chartjunk with ``graphical decoration", Tufte muses that it is included by a designer ``to make the graphic appear more scientific and precise, to enliven the display, to give the designer an opportunity to exercise artistic skills." He argues for the expulsion of chartjunk from the pursuit of clear, truth-seeking charts, and characterizes the work to produce chartjunk as misplaced effort: ``Graphical decoration, which prospers in technical publications as well as in commercial and media graphics, comes cheaper than the hard work required to produce intriguing numbers and secure evidence"\cite[p76,107]{tufte2001}.

Characterizing chartjunk as all the ink in a chart that is not \textit{data-ink} -- ``non-erasable" ink used in a (printed) chart that is ``is arranged in response to variation in the numbers presented" -- Tufte provided a mechanism to quantify the ``proportion of a graphic that can be erased without loss of data-information" through what he called the \textit{data-ink-ratio}. This ratio provides a way to approximate the amount of chartjunk in a visualization, and a path towards Tufte's minimalist goal: ``the larger the share of a graphic's ink devoted to data, the better (other relevant matters being equal)"\cite[p93,96]{tufte2001}. 

Although Tufte grounds his chartjunk provocation in a critique of designerly approaches to visualization design, he applies his ideas of reducing non- and redundant-data-ink to a broader class of graphic elements. ``Fortunately most chartjunk does not involve artistic considerations. It is simply conventional graphical paraphernalia routinely added to every day display that passes by: over-busy gridlines and excess ticks, redundant representations of the simplest data, the debris of computer plotting, and many of the devices generating design variation"\cite[p107]{tufte2001}. Through an extreme position of minimal design, some of Tufte's best known redesigns eliminate otherwise standard graphical elements like axis-lines and the left-vertical line of bars, a view that others have critiqued as overly difficult to interpret\cite{bateman2010, chabris2005}.

\subsection{Evolution}

With the success of \textit{The Visual Display of Quantitative Information} among visualization practitioners and researchers who embraced its message of minimalism, Tufte's chartjunk provocation became embedded into the lexicon of visualization design. There, chartjunk became an umbrella term to describe a broad range of graphical devices used by practitioners and studied by researchers. In contrast to Tufte's original extreme-minimalism definition of chartjunk, practitioners and researchers have taken more nuanced interpretations of what constitutes junk in a chart.

Steven Few, a well-known practitioner and book author who also encourages minimal visualization design, says that he agrees with Tufte's assessment that chartjunk contains non-useful information, but differs in what he would consider to be ``useful expressions of information"\cite{few2011}. ``Embellishments can at times, when properly chosen and designed, represent information redundantly in useful ways, and even when they aren't information in and of themselves, can meaningfully support the display of information"\cite{few2011}. He calls out Tufte's original definition of chartjunk as ``too loose", and acknowledges the provocative origins of the term: ``By defining chartjunk too broadly, Tufte to some degree invited the heated controversy that has raged ever since."

Robert Kosara similarly argues in a blog post that not all chartjunk is bad\cite{kosara2012}, contending that some types of chartjunk are harmless, such as a ``pretty picture” or an ``elaborate border”, while others can even be helpful like annotations and explanatory text. He proposes new adjectives to help differentiate between good and bad non-data-ink: \textit{useful} junk, \textit{harmless} junk, and \textit{harmful} junk. 
Pointing out the lingering negative connotation of these terms, commenter Roman pushes back on this proposal in the comments: ``If junk is useful, then it's not junk anymore."

In a recent interview study, Parsons \& Shukla explored the varied perspectives of practitioners on their use of chartjunk\cite{parsons2020}. The participants noted the inherent negative connotation associated with the term, as well as a current ``corrective movement" away from Tufte’s extreme-minimalism style to one that makes space for more personal and situated considerations like aesthetics, style, context, and client constraints. They diverged, however, in what they would label as chartjunk based on their personal preferences: ``While the vast majority of participants indicated that embellishments had a place as long as they fit the context and other personal and situational goals, participants would regularly reveal underlying commitments to a particular design philosophy when justifying their views, often surfacing value-laden judgments about the proper ways in which visualizations should be designed and consumed." The pluralistic views of the participants revealed chartjunk as a proxy for bad design, as opposed to a precise term for describing graphical devices. The authors offer that one conclusion of their research is that ``better definitions are needed so that everyone has a shared understanding."

Several papers reporting on empirical studies that look at the effects of chartjunk on viewers' experiences incorporate a range of additional terms in an attempt to more precisely describe the specific graphic devices under study: 
embellishment\cite{bateman2010,wu2021},
excessive annotation\cite{borkin2013},
decoration\cite{borkin2013},
extraneous elements\cite{hullman2011},
and unnecessary ornamentation\cite{hullman2011}. A close reading of these papers also reveals the lingering effects of Tufte's provocation on the studies. A study by Bateman et al. on the memorability of visualizations containing chartjunk, compared charts designed by Holmes with unadorned, ``plain" charts designed by the authors, finding that the Holmes charts were more memorable\cite{bateman2010}. A critique of the study by Few contends that the plain charts -- which he called ``plain ugly" -- did not abide by Tufte's minimalist standards, nor the design standards of the day, bringing the generalizability of the findings into question\cite{few2011}. A follow-up study by Borkin et al. took a different approach by studying the memorability of a significantly broader range of charts found in the wild \cite{borkin2013}. In this study the charts were categorized based on their visual density -- a proxy metric for the amount of chartjunk each contained. Charts with low visual density, and thus low amounts of chartjunk, were labeled ``good", and those with high visual density labeled ``bad". The results of the study indicate that good charts are bad for memorability, and some bad charts are good, producing a dissonance between what is good and what is bad. This dissonance also appears in a recent study by Wu et al. that found chartjunk to provide accessibility benefits to people with intellectual disabilities \cite{wu2021}.

The studies we cite here are just a few of the many that have attempted to provide empirical evidence to fill the void left by Tufte who proclaimed, without any proof~\cite{inbar2007}, that chartjunk is bad and data-ink is good.
The combined results of these studies, however, are contradictory and inconclusive, leading some researchers to conclude that in the end, these aren't the right guidelines for reliably designing visualizations that work for everyone in every context in the same way~\cite{gillan1994,correll2014,kostelnick2008,hill2016}.

\subsection{Dissolution}

Our telling of the origin and evolution of the term \textit{chartjunk} is a provocation to acknowledge its history and its baggage, and to ask the question: where do we go from here?
Do we continue to fumble for words that attempt to neutralize the negative connotations of junk?
Do we continue to devalue designerly and other non-minimalist perspectives of good visualization design?
Do we continue to produce incongruities in how we write about, talk about, and think about what makes visualizations effective?

Chartjunk is a provocation. It is not precise, dispassionate, or impartial, nor was it meant to be. And yet, as researchers we have taken the term seriously. We call for the dissolution of the term from our research lexicon for referring to the myriad of graphic devices that make up a visualization. What to replace chartjunk with is a question that goes beyond the scope of this paper. What we do offer, however, is a performance of maintenance, where we take the first steps to acknowledge, remove, and re-imagine chartjunk.

\section{Performing maintenance art on chartjunk}

We recognize that it is common for visualization authors to offer specific and actionable solutions. However, we cannot yet offer replacement terms for chartjunk given the breadth of graphical devices it was (and still is) meant to describe. Instead, we offer a performance of maintenance that works \textit{towards} removing chartjunk from our research lexicon; a taking out of our community’s trash, in three parts. First, by editing the Wikipedia page on chartjunk to include our more complete etymology of the term: an act of acknowledging chart junk’s provocative origins and ill-fated evolution. Second, by physically cutting out chartjunk from pages of IEEE research articles: an act of removing chartjunk from our lexicon. And third, by scanning and posting a repository of the pages with chartjunk removed: an act of inviting the community to re-imagine what terms we might use instead.

Our performance is inspired by the work of art activist Mierle Laderman Ukeles who responded with a crafting of maintenance art in the 1970s when her role as a mother subsumed her identity as artist. She was the first artist-in-residence at NYC's Department of Sanitation and focused on depicting maintenance work as art. Her art(work) was tedious, like sweeping the streets of NYC as performance art, and shaking the hand of every garbage(wo)man, but it worked to highlight the necessity of maintenance and the invisible labor behind upkeep. It was through the act of maintenance, conceived as performance art, that Ukeles reframed the conversation around maintenance work. We, too, rely on maintenance art to reframe the conversation surrounding chartjunk, contributing our own performance of maintenance work toward changing our community's use of the term.

And though these actions were ephemeral, completed as part of a performance, we present a triptych of artifacts that are the traces\cite{rogers2020, offenhuber2019} of our maintenance work: The Edit, The Video, and The Scanned Repository. These artifacts span the past, present and future. They are symbols of maintenance work --- reminders of work that has been done, visual representations of our labor, and persistent invitations to continue maintenance work into the future.
Despite the artifacts spanning different mediums, each piece of the triptych is united by a dedication to playfully blurring the boundaries between research, activism, and maintenance art.

\begin{figure}[h]
    \centering
    \includegraphics[width=8cm]{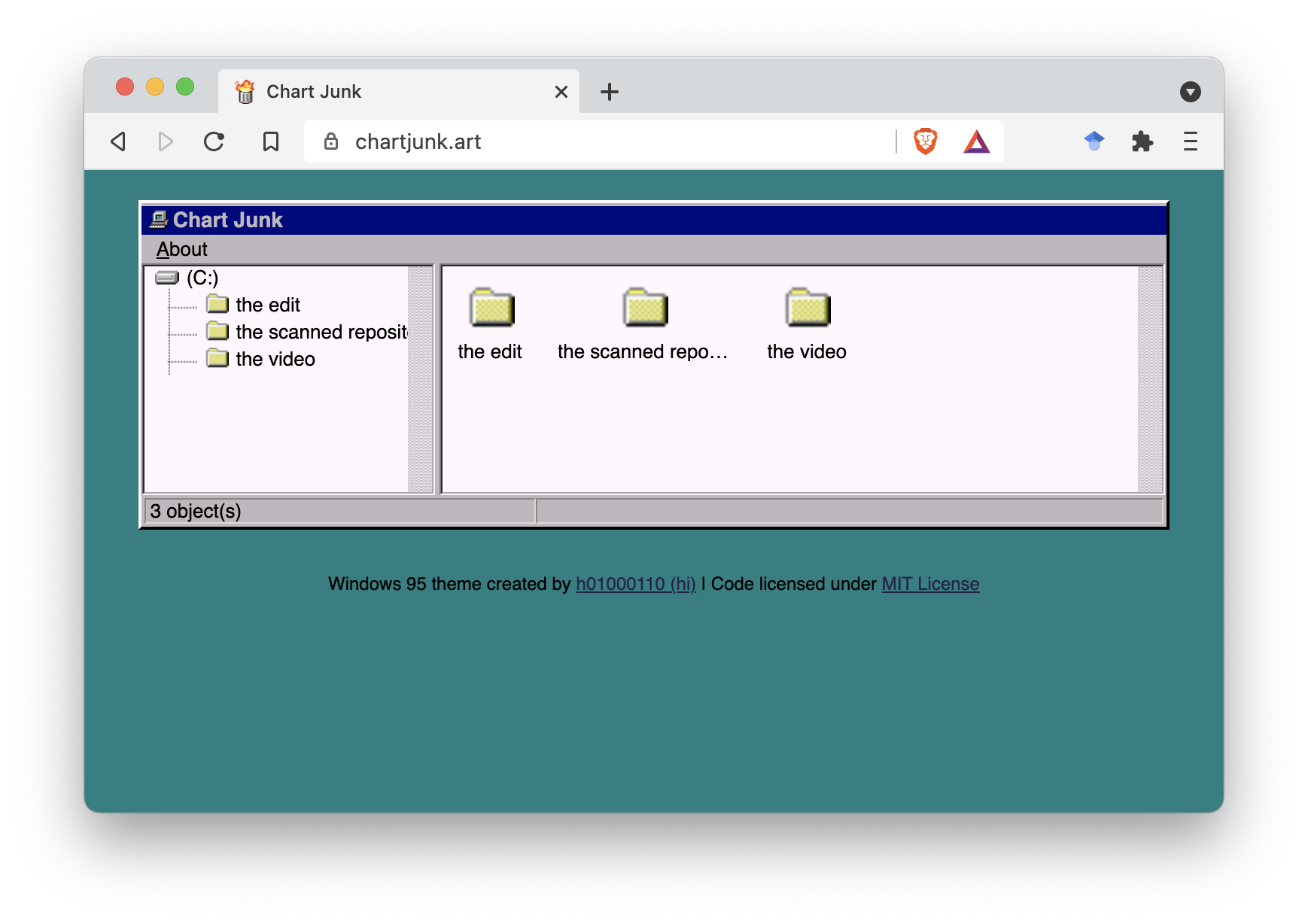}
    \caption{The website \url{https://jackwilb.github.io/chart-junk/} hosts the triptych of maintenance performance artifacts. Each folder contains a ReadMe document with more information about the artifact and the artifact itself. We invite readers to explore and engage with the artifacts, and perhaps suggest their own maintenance to the site.}
    \label{fig:website}
\end{figure}

The artifacts are stored online at \url{https://jackwilb.github.io/chart-junk/}, using an intentional aesthetic. Shown in Figure \ref{fig:website}, the aesthetic is a reincarnation of the Windows 95 desktop, complete with chunky borders, 3D-stylized buttons, and pixelated icons. Although the work is focused on maintenance, which implies an orderliness or cleanliness, we specifically chose to stay away from minimalist design since the work of maintenance is never finished. Even our use of a flaming garbage can as a favicon is an act of centering humor to evoke our call to take out the trash, a gentle reminder that we, the research community, have taken the chartjunk provocation too seriously. We invite you to explore the website as you read through the rest of this section, which details our three enactments of maintenance and their resulting artifacts. We offer these actions, exaggerated under the guise of performance, to the community as first steps in removing chartjunk from our lexicon. Our intention is to inspire others to join us in maintenance art.

\subsection{The Edit}
\textit{\url{https://jackwilb.github.io/chart-junk/tag/the-edit/}}

For our first performance of maintenance, we edited the Wikipedia entry on chartjunk.  This act of maintenance \textit{acknowledges} the provocative origins of chartjunk and its uptake by our community and results in a digital artifact of Wikipedia edits. We see the performance of maintenance akin to renovating an old house; the resulting house has the same address, but the updates increase functionality and change how the house is perceived. We felt that it was important to edit the Wikipedia page for a few reasons. 
First, by performing maintenance on the Wikipedia entry, we are publicly reframing chartjunk as a provocation instead of a technical term to be taken seriously.
Second, we believe that as a research community, we can, and should, have a more active role in making visualization research accessible to non-researchers. 
And last, there is a large disparity across Wikipedians, those who edit Wikipedia, along gender lines \cite{wikipedia2021}. 
As part of our dedication to maintenance art, a performance that makes hidden labor visible, the first author who identifies as a woman, found it important to contribute to a knowledge repository that is currently dominated by men.

Prior to our edits, the Wikipedia entry contained only a couple paragraphs that briefly reviewed the definition of chartjunk provided by Tufte. Our edits add richer context and a diversity of sources to the entry. Specifically, we added two new sections to the Wikipedia article: etymology and the focus of chartjunk in research.
% On July 30, 2021, we edited the Wikipedia page to include two new sections: etymology and focus in research. 
By including the etymology of the word, our intent is to ground the extant definition of the term, currently present in Wikipedia, in its historical context as a provocation by Tufte. Explicating the history of a term can help contextualize and situate it for more nuanced discussions relevant to present topics, as seen in the \textit{The Atlantic}'s article about the etymology of infrastructure \cite{atlantic2021}. We also added a synopsis of research findings that demonstrate the potential benefits of chartjunk. Again our intent here is to complicate the narrative that chartjunk is only bad. 

The resulting artifact, which we have called The Edit, is the first step in maintenance. The Wikipedia entry is as an open invitation to other members of the community to further contribute to the maintenance of chartjunk by adding their edits or disputing ours. We also hope that our work engages members outside of our community to make their own edits. Through active maintenance of this Wikipedia page, we see the potential to record and expand conversations started in this paper. 

\subsection{The Video}
\textit{\url{https://jackwilb.github.io/chart-junk/tag/the-video/}}

For our second performance of maintenance, we focused on \textit{removing} chartjunk, through cutting printed text of IEEE papers that used the term. This act of maintenance is akin to taking out the trash. We felt the need to remove the printed words from the papers to make a statement about the reliance our community has had on using the term as a stopgap. We performed the act of removal and uploaded the video with very few edits to authentically demonstrate the time and resources it took to physically remove chartjunk. 

This performance of removal as maintenance is boring. As  Ukeles' admits in \textit{Manifesto for Maintenance Art 1969!}, maintenance is boring and ubiquitous: ``Maintenance is a drag; it takes all the fucking time (lit.)'' \cite{ukeles2018}. And as part of this manifesto, she proposes a museum exhibit where everyday she would clean the museum: ``sweep and wax the floors, dust everything, wash everything.'' Her frankness is refreshing. For at the core of maintenance art is a recognition that sometimes we need to do boring things, but that a focus on these actions is a powerful mechanism for revealing ignored and invisible labor. 

Our video is an homage to Ukeles' performance art. In the video, the first author is tediously cutting out the term chartjunk from all IEEE papers that use the term. To create the corpus of papers, we used the following search terms in IEEE Xplore:
\begin{enumerate}[nosep]
    \item (``Full Text \& Metadata'':``chart junk'') OR (``Full Text \& Metadata'':chartjunk) OR (``Document Title'':``chart junk'') OR (``Document Title'':chartjunk)
    \item Filters Applied: Journals
\end{enumerate}
This query produced 28 results, from which we removed 4 because they were book reviews. For printing and cutting the papers, we collated all the documents and then ran an Adobe script to highlight and extract pages with the terms $\{chart junk, chart-junk, chartjunk\}$. We did this so that we would not waste more paper than necessary. The Adobe script resulted in 84 extracted pages and these pages are the pages that the first author cut.

The physical act of cutting out a term from research papers took time, a precious resource for a graduate student, and the video captures the cost associated with performing this maintenance. While a few hours of preparation, performance, and video editing were worth it for the production of a performative piece, we want the video to serve as a reminder of the consequences of using ill-defined words. We invite readers to question: who is responsible for maintenance? If it weren't for the performance and creation of a video, how could we attribute recognition to maintenance work?

\subsection{The Scanned Repository}
\textit{\url{https://jackwilb.github.io/chart-junk/tag/the-scanned-repository/}}

For our final performance of maintenance, we scanned and posted pages from research papers missing the word chartjunk. This final piece of our triptych is an invitation to \textit{re-imagine} chartjunk. We present a scanned repository as a discussion starter, where blank spaces are opportunities for our community to begin the maintenance work of brainstorming different terminology to replace chartjunk in our research lexicon. The invitation is also a request to share the burden of maintenance. Unlike the other maintenance performances, we see this work as ongoing, distributed, and generative, like picking up trash along a beach's shoreline.

While cutting out the term chartjunk from papers, the first author was careful to cut in-between the lines so that the resulting pages could remain legible. The resulting pages are intact except for the areas where the term had been cut out, resulting in a Mad Libs-esk document. The pages are accessible through the website, each one separated into individual pages. We chose to separate papers into individual pages to digitally replicate the first author's embodied experience of picking up and cutting each page individually. We also separate the pages from papers as respect to the authors of those papers -- by not grouping pages from the same paper, and by not including the entirety of the paper, we wish to demonstrate that our maintenance performance is on the term chartjunk and not on the authors' research.

Each page is available online and we invite readers to look through them. We see the blank spaces as a collective forum for people to engage with and debate new terms that may fill these gaps. 
We’ve created an online avatar on Twitter, named Main T. Nance (\href{https://twitter.com/SayNo2Chartjunk}{@SayNo2Chartjunk}), that can be tweeted at with these new alternative terms; this avatar is a co-author of this paper. 
We invite readers to tweet and perform their own acts of maintenance. Tweeting at this new avatar can be easily accomplished by clicking the tweet button on any page and providing your suggestion. This display of the pages and the additional artifacts, generated as tweets, will remain as a testament to our community's commitment to maintenance.

\section{Conclusion}

\begin{quote}
    ``After the revolution, who’s going to pick up the garbage on Monday morning?" - \textit{Mierle L. Ukeles}
\end{quote}

The work of maintenance is never complete. Chartjunk began as a provocation by Tufte, to dismiss all but the most minimalist visual elements in visualization design. And his provocation worked. We took the term seriously, resulting in both devaluing of designerly ways of knowing and fumbling for years with an imprecise term. But now it is time to take the steps toward removing chartjunk from our research lexicon. Towards this goal we playfully performed maintenance art on chartjunk: maintenance as art and art as maintenance. The traces of our performance, presented as a triptych of artifacts, act as visual reminders of our work and open invitations to the community to join in on maintenance (art)work.

One final question remains: where do we go from here? After throwing out chartjunk we leave a hole behind, both physical -- in the form of cut up pages -- and epistemic -- in the form of missing vernacular to describe graphical devices. We could choose to adopt new terms to more accurately describe the minimalist design principles, designing charts similar to modern guidelines, just using more precise terminology; we could choose to modify our design ideologies and focus on the aspects of design that enhance a user's experience with charts, completely reshaping our design principles from the ground up; or we could take a moderate option somewhere between the two. In any case, let this manifesto be a turning point in the chartjunk debate and a catalyst for change.

% \input{sections/conclusion}

\acknowledgments{
The authors wish to thank the Visualization Design Lab for inspiring conversations that sparked the genesis of this paper, Mierle Laderman Ukeles for maintenance art and art maintenance, and for all of the people out there dedicated to maintenance work, even when it is a drag. This work is partially funded by the National Science Foundation (OAC 1835904), and by the Wallenberg AI, Autonomous Systems and Software Program (WASP) funded by the Knut and Alice Wallenberg Foundation.}

\bibliographystyle{abbrv-doi}

\bibliography{main}
\end{document}